\documentclass[useAMS,usenatbib,usegraphicx]{mn2e}
\newcommand{\km}{km~s$^{-1}\;$}
\newcommand{\kms}{km~s$^{-1}\;$}
\newcommand{\kmss}{km~s$^{-1}$}

\newcommand{\hi}{H{\sc I}$\;$}
\newcommand{\hii}{H{\sc I}}
\newcommand{\vlsr}{V$_{\rm LSR}\;$}

\newcommand{\mb}{mJy  beam$^{-1}$}
\newcommand{\jb}{Jy  beam$^{-1}$}

\newcommand{\ho}{H$_{2}$O$\;$}

\title[]{VLBI imaging of OH absorption: The puzzle of the nuclear region of 
NGC\,3079}
\author[]{Yoshiaki Hagiwara,$^{1}$\thanks{E-mail: hagiwara@astron.nl (YH)} 
Hans-Rainer Kl\"ockner,$^{2,1}$ Willem Baan$^{1}$ \\ 
$^{1}$ASTRON, Westerbork Observatory, P.O. Box 2, Dwingeloo, 7990 AA, The 
Netherlands\\
$^{2}$Kapteyn Astronomical Institute, University of Groningen, Postbus 800, 
Groningen, The Netherlands}
\begin{document}
\date{2003}


\maketitle

\label{firstpage}

\begin{abstract}
Broad hydroxyl (OH) absorption-lines in the 1667 MHz and 1665 MHz
transition towards the central region of NGC\,3079 have been observed
at high resolution with the European VLBI Network (EVN). Velocity
fields of two OH absorption components were resolved across the
unresolved nuclear radio continuum of $\sim$ 10 parsecs.  The velocity
field of the OH absorption close to the systemic velocity shows
rotation in nearly the same sense as the edge-on galactic-scale
molecular disk probed by CO(1--0) emission. The velocity field of the
blue-shifted OH absorption displays a gradient in almost the opposite
direction.  The blue-shifted velocity field represents a
non-rotational component, which may trace an outflow from the nucleus,
or material driven and shocked by the kiloparsec-scale
superbubble. This OH absorption component traces a structure that does
not support a counter-rotating disk suggested on the basis of the
neutral hydrogen absorption.
\end{abstract}

%
%
%
%
\section[]{Introduction}
NGC\,3079 is an edge-on Sc galaxy housing a LINER (Low Ionisation
Narrow Emission Line Region) nucleus \citep{heck80} with a long
history of observations at various wavelengths. The nucleus is also
classified as a type 2 Seyfert nucleus \citep{ford86}. The galaxy has
a systemic velocity of 1116 \km \citep{irwi91}, which gives the
distance of 16 Mpc and hence 1 milliarcsec (mas) corresponds to
approximately 0.08 pc, adopting H$_0$ = 75 \kms
Mpc$^{-1}$. Observations of optical emissions with the Hubble Space
Telescope (HST) have shown a number of outflowing filaments in a
'superbubble', produced by a combination of stellar winds and
supernova explosions in a site of massive star formation.  X-ray
observations with Chandra show a clear correspondence of
H$\alpha$-line filaments with those in X-rays at the distance of $\sim
$ 65 pc from the nucleus \citep{ceci02}. Throughout several X-ray
observations, a highly obscured active galactic nucleus (AGN) and iron
line emission towards the nucleus were found \citep{ceci02}. The
center of the galaxy has been resolved by radio interferometry and
reveals a double radio lobe extending 1.5 kpc on either side of the
major axis of the galaxy and a puzzling nuclear structure on parsec
scales \citep{irwi88}. Due to the complexity of the radio structure,
the interpretation of the location of a true nucleus is not
straightforward.  \\ 
Broad (200--400 \kmss) and deep ($\tau$ $\sim$
0.5) neutral hydrogen \hi absorption and weaker but equally
broad and deep OH absorption are found towards those compact radio
sources in the nucleus. These absorptions have been studied in detail
at various angular resolutions by radio interferometers.
Based on the Very Large Array (VLA) observations at 1 arcsec resolution, 
\cite{baan95} considered that the two components of \hi and OH absorptions are
associated with an obscuring torus in the nuclear
region. Multi-Element Radio Linked Interferometer Network (MERLIN)
observations at sub-arcsecond resolution resolved the rotation of the
\hi absorption \citep{pedl96} in the same sense as the rotational
trend traced by \hi emission and CO(1--0) emission on galactic scales
\citep{irwi91,irwi92}. According to Very Long Baseline Interferometry
(VLBI) observations at an angular resolution of 10-15 mas, three
\hi absorption components viewed against the resolved double-peaked
radio source appear to be evidence for counter rotation relative to 
the rotation in the outer galaxy \citep{sato00}.  \\
The strong and highly intensity-variable water maser emission is known
to exist in NGC\,3079 \citep{henk84,hasc85}. Earlier attempts for
resolving the maser emission have been made since the late 1980s using
VLBI \citep{hasc90}. At present, it is understood that the \ho maser
components are distributed within an elongated (disk-like) structure
within a few parsecs from the nucleus \citep{trot98}.  Most of the
maser components detected with VLBI lie in a velocity range \vlsr
(LSR refers Local Standard of Rest.) = 956--1190 \kms and are not
associated with any of the jet components \citep{trot98,sato00}. They
are distributed nearly north-to-south, and are aligned roughly with
the molecular disk traced by CO(1--0) (Position Angle (P.A.) =
15$\degr$) with a radius of 750 pc \citep{irwi92}.  Several new weak
components were recently detected, which extend the velocity range to
about \vlsr = 1350 \kms \citep{hagi02a}. Given the velocity
distribution of all the maser features observed to date, the maser
spectrum shows a rough symmetry w.r.t the systemic velocity of the
galaxy (\vlsr $\simeq$ 1116 \kmss) \citep{hagi02b}.  It has been
argued that there is a rotating parsec-scale molecular gas disk or
torus obscuring an AGN in NGC\,3079, but no concrete evidence has been
presented in the literature like the one in NGC\,4258
\citep{herr98}. A recent study by \cite{enno03} presents evidence for
five distinct radio continuum sources in the nucleus of NGC\,3079,
complicating the interpretation of its puzzling structure. Besides the
three components A, B, and C that are detected at higher frequencies
and are seen at 22 GHz together with the \ho masers, there are two
other components E and F that are visible only at lower frequencies,
following the convention used in \cite{irwi88}. These additional
components have the same orientation and are equally spaced as the 22
GHz components A and B, but they are shifted about 25 mas $\simeq$ 2
pc to the east; they form the 18-cm radio signature that is observed
in \hii\ and OH absorption experiments.\\ 
In this paper, we present
VLBI observations of OH absorption at the nuclear region of
NGC\,3079. With observations using European VLBI Network (EVN) at the
highest spectral-line sensitivity of any existing VLBI facility, we
aim to review the kinematics of the circumnuclear region in the
galaxy.\\
\begin{table}
\caption{Properties of NGC 3079. Coordinates in this table were phase
centre values used for the data correlation at JIVE. Velocity with
respect to LSR for radio definition is adopted from Irwin \& Seaquist
(1991). Optical classification was made by Heckman (1980) and Ford et
al. (1986). Radio flux densities were measured by the VLBA
(Sawada-Satoh et al. 2000). Infra-red (IR) luminosity is from Baan
(1989). X-ray luminosity is from \citet{ceci02}. We assume that $H_0$
= 75 km s$^{-1}$ Mpc$^{\;-1}$.}
  \begin{tabular}{@{}ll@{}}
  \hline 

RA   (J2000)  &  10$^{\rm h}$01$^{\rm m}$57.8050$^{\rm s}$ \\
Decl (J2000)  &  55$\degr$40'47.080'' \\
Systemic Velocity (21-cm HI)  & 1116 $\pm$ 1  km s$^{-1}$  \\
Distance                &  16 Mpc \\
Inclination (optical)  & 84$\degr$ \\
Optical class    & LINER/Seyfert 2\\
F$_{\nu}$ (1.4 GHz) & 5.7 $\pm$ 1.7  mJy \\
~~~~~(8.4 GHz) & 14.2 $\pm$ 0.7 mJy \\
~~~~~(15  GHz) & 24.4 $\pm$ 0.6  mJy \\
~~~~~(22  GHz) & 12.1 $\pm$ 0.7 mJy \\
L$_{\rm IR}$  & 3 10$^{10}$ erg s$^{-1}$  \\
L$_{\rm X-ray (0.1-6.5 keV)}$  &  1.0 10$^{42}$ erg s$^{-1}$ \\   
L$_{\rm X-ray (Fe~line;6.4 keV)}$  &  3.3 10$^{41}$ erg s$^{-1}$ \\
\hline
\end{tabular}
\end{table}            
%
%
%
%
\section[]{Observations}
The central region of NGC\,3079 was observed at 18-cm on 2000 November 15 with 
the EVN, which consisted of eight telescopes: Cambridge, Effelsberg, 
Jodrell Bank (Lovell), Medicina, Noto, Onsala, Torun, and the Westerbork 
phased-array. We observed the 1667 and 1665 MHz main line transitions of the 
$^2\prod_{3/2}$, J=3/2 ground-state of OH in absorption toward the central 
radio continuum source in NGC\,3079. \\
NGC\,3079 was observed in a phase-referencing mode over a period of
5.7 hrs, interspersed with observations of a phase calibrator source
J\,0957+5522, which is located about 1 degree from the target
source. The total time of one observing cycle was 13 minutes with 10
minutes for NGC\,3079 and 3 minutes for J\,0957+5522. The total time
spent on the target was about 3 hours. The data were recorded in both
left and right circular polarizations using a single intermediate
frequency (IF) band with 8 MHz bandwidth. The IF baseband was
subdivided into 128 spectral points, yielding a frequency resolution
of 62.5 kHz, or 11.2 \kms in velocity at the source distance. The IF
velocity coverage is 1440.5 \kmss. In order to cover both the main
line transitions, we centered the IF on \vlsr = 1350 \kms and the
Doppler velocity center was referenced to the 1667 MHz
line. Hereafter, all velocities are in the radio convention and with
respect to LSR.  The system temperature and antenna sensitivity for
each EVN antenna over the observing frequency range from 30 K $-$ 100
K during the observing run.  DA\,193 was measured in the middle of the
observations for calibrating the absolute amplitude gain and the
bandpass correction. The data-recording rate was 128 Mbit per second
with a MkIV terminal. \\

The correlation of the data was performed at the EVN MkIV correlator at the 
Joint Institute for VLBI in Europe (JIVE). Data analysis was made using the 
NRAO AIPS package. After the delay and delay rate calibration using 
J\,0957+5522, the bandpass corrections were applied. We discarded the data from six baselines,  Noto(NT)-Medicina(Mc), Mc-Torun(Tr), Tr-NT, Cambridge(Cm)-Tr, Cm-NT, and Cm-Mc baseline, due to inadequate data quality. The continuum visibility data set was 
generated from the spectral-line visibility data set by averaging 
absorption-free channels using the AIPS task of UVLIN. The continuum visibility 
data set was then used for self-calibration to improve the image sensitivity. 
After iterations of the self-calibration in both phase and amplitude, the 
solutions were transferred to the spectral-line visibility data. The continuum 
and spectral-line visibilities were imaged employing IMAGR.


\section[]{Results}
 Fig. 1 ($left$) shows the EVN spectrum of OH absorption integrated
over the nuclear radio continuum source (Fig. 4) with natural weighting
with a spectral resolution of 11.2 \kmss. The systemic velocity of the
galaxy \vlsr = 1116 \kms and the peak velocities of HI absorption and
water maser emission are denoted. The rms noise level is 0.7 \mb\ per
channel. The OH main-line transitions of 1667 MHz and 1665 MHz are
clearly detected against an unresolved continuum source at the centre
of NGC\,3079. Both transitions have two distinct components, one of
which is blue-shifted (OH1) and the other is nearly centred on the
systemic velocity (OH2). This is consistent with the spectra at lower
resolutions \citep{baan95}. By comparison with the single-dish
spectrum, the OH absorption lines observed in these observations have
lost their broad wings. The integrated intensities of the 1667 MHz OH
absorption estimated from the single-dish and the EVN spectrum in
Fig. 1 ($left$) are about 8.3 Jy \kms \citep{baan95} and 0.74 Jy
\kmss, respectively. Only 8.9 percent of the absorption intensity was
recovered in the EVN observations. This accounts for the narrower
line-widths of the OH absorption spectra obtained with the EVN (Table
2). The OH emissions in the wing of the absorption at \vlsr = 910 and
around \vlsr = 1260 \kms suggested by \cite{baan95} coincide with weak
emission features seen in our EVN spectrum, although both of them need
to be confirmed in further observations. Two Gaussian components have
been fitted to the absorption profiles of each of the transition
(Fig. 1 $right$); the results are listed in Table 2. The resultant
opacity and column density are listed in Table 3 together with those
of the previous VLA and VLBI observations obtained at different
angular scales. The OH absorption velocity center (Gaussian-fitted) is
\vlsr(OH1)= 1011.9 $\pm$ 0.9 \kms and \vlsr(OH2)= 1113.5 $\pm$ 2.0
\kmss, while the centre velocities derived from the VLA-A data are
\vlsr = 1013 \kms and \vlsr = 1114 \kms \citep{baan95}. These values
are consistent within the spectral resolution of one channel. The HI
absorptions are peaked at \vlsr = 1010 \kmss, 1120 \kmss, and 1265
\kms \citep{baan95}, where the red-shifted third component has no OH
counterpart. The ratios of the double peaks in both transitions are
quite consistent, which may imply that both main-line features are
associated with the same continuum source. The line ratio 1667 to 1665
MHz in the Gaussian-fitted profiles is 1.5, which is lower than the
Local Thermodynamic Equilibrium (LTE) value of 1.8.\\
Velocity-integrated maps of the 1667 MHz OH absorption integrated over
11.2 \kms intervals and the CLEAN map of a 18-cm radio continuum
source are shown in Fig. 2 (All are uniformly weighted.). Fifteen
velocity channel maps cover the velocity range of the absorption from
990.4 \kms to 1147.7 \kmss. \\
The continuum emission was not resolved in
this experiment. The peak flux density of the continuum map produced
with uniform weighting is 8.8 \mb\ and the rms noise level is 0.25
\mb. The peak flux density and the noise level of the continuum map
produced with natural weighting are 11.8 \mb\ and 0.085 \mb,
respectively.  The integrated flux density of the continuum is 14.3
$\pm$ 0.7 mJy, while the 21-cm integrated flux density obtained by
single-dish measurements is 760 $\pm$ 31 mJy \citep{cond83}. Hence,
more than 95 percent of the continuum flux is missing in this EVN
experiment.  Because the limited angular resolution of our EVN
observations compared with the past VLBI observations at higher
frequencies, the nuclear continuum remains unresolved without showing
any structure. The EVN synthesized beam is larger than the angular
separation between the various nuclear components (Fig. 7), which
limits the discussion on the possible associations of the OH absorbing
gas with individual sources in later sections. \\
A position-velocity (PV) diagram of the 1667 MHz absorption is
presented in Fig. 3. A cut was made through the continuum emission at
P.A. = 15$\degr$ (rotated clockwise by 15 degrees), which aligns a cut
along a major axis of the rotation traced by CO(1--0)
\citep{sofu92,koda02} and which was also used from the PV diagrams in
\cite{baan95}.  The gradients in the 1665 MHz line show the same trend
as those seen in the 1667 MHz line. Note that there are significant
position offsets of $\sim$7-8 mas in declination between the two peaks
of OH1 (blue-shifted) and OH2 (systemic), which could not be measured
in the PV diagrams of \cite{baan95} at 1 arcsec resolution with the
VLA-A. The velocity contours of OH1 and OH2 are separate, as compared with the PV diagrams in Fig. 7 in \cite{baan95}.
\\
Fig. 4 shows a naturally weighted 18-cm continuum map superposed on
the integrated 1667 MHz OH intensity (0th moment) map. The absorption
is apparently concentrated towards the centre with a weak outward
extension. This is similar to the east-west
elongation of the \hi and OH absorbing gas seen at 1 arcsec resolution
and caused by the orientation and separation ($\sim$ 25 mas or $\sim$
2 pc) of the L-band components E and F \citep{irwi88, baan95, enno03}. \\
Fig. 5 displays the OH velocity field of the two components (1st
moment). The velocity field of OH1 varies from southwest to northwest
in P.A.= 60$\degr$, while that of OH2 is seen approximately from north
to south in P.A.= 145$\degr$. The velocity gradients of OH1 and OH2
are $\sim$ 10 \kms pc$^{-1}$ and $\sim$ 13 \kms pc$^{-1}$,
respectively. The kpc-scale CO(1--0) velocity gradient is found to be
also in the north-to-south direction and has a value of 0.85 \kms
pc$^{-1}$ and a P.A.= 15$\degr$ \citep{sofu92}. The compact nuclear 
CO(1--0) core at PA= 4$\degr$, tilted by about 10$\degr$ w.r.t the kpc-scale
CO(1--0) disk, within central 125 pc shows a rigid rotation with a
velocity of $\sim$ 300 \kms \citep{sofu01, koda02}. This yields the
velocity gradient of 2.4 \kms pc$^{-1}$. Just as \cite{baan95} also
\cite{sato00} have interpreted the equivalent \hi components to
originate against the two continuum components A and B, which in
reality were components E and F \citep{enno03}. The velocity
differences estimated from the CO velocity gradients between the two
continuum components F (A) and E (B), separated by 25 mas or 2 pc, are
only 1.7 \kms or 4.8 \kmss. Such values are not
consistent with the observed velocity difference of 101.6 \kms between
OH1 and OH2, which would suggest that the two detected OH absorptions
arise from neither the kpc-scale CO disk nor the nuclear CO
core. MERLIN observations at $\sim$ 400 mas resolution resolved the
rotational motion of the \hi absorbing gas \citep{pedl96}, which
matches the sense of the rotation traced by the \hi emission
\citep{irwi91}. The directions of the \hi velocity fields in the three
\hi VLBI absorption components have not been presented in
\cite{sato00}, which prevents a comparison with the OH velocity
fields in Fig. 5. \\
Comparisons of the intensity distribution of OH1 and OH2 for each 
transition are presented in Fig. 6. The intensity maps were produced from the 
uniformly weighted spectral-line cubes. We find that there is a difference in 
the spatial distribution in both 1667 MHz and 1665 MHz absorption, which is 
even more distinct in the 1665 MHz line. The distribution of the blue-shifted 
OH1 is downward w.r.t that of the systemic OH2 by approximately 8 mas, or 0.6 
pc in declination. This trend is also true in the PV diagram of Fig. 3, where 
the difference is clearly seen for both main-lines. \\
\begin{figure}
\includegraphics[angle=0,scale=0.45]{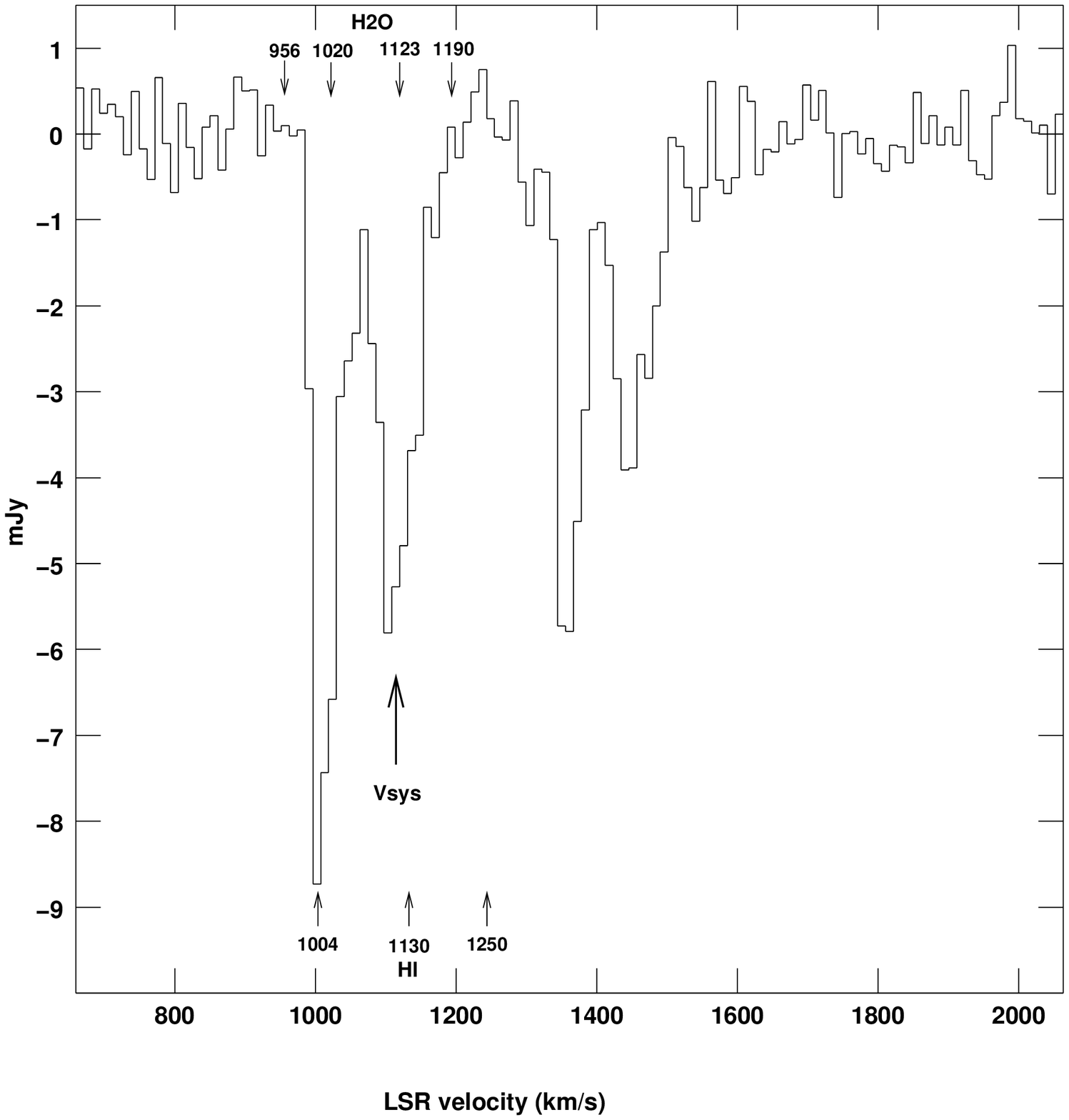}
\includegraphics[angle=0,scale=0.52]{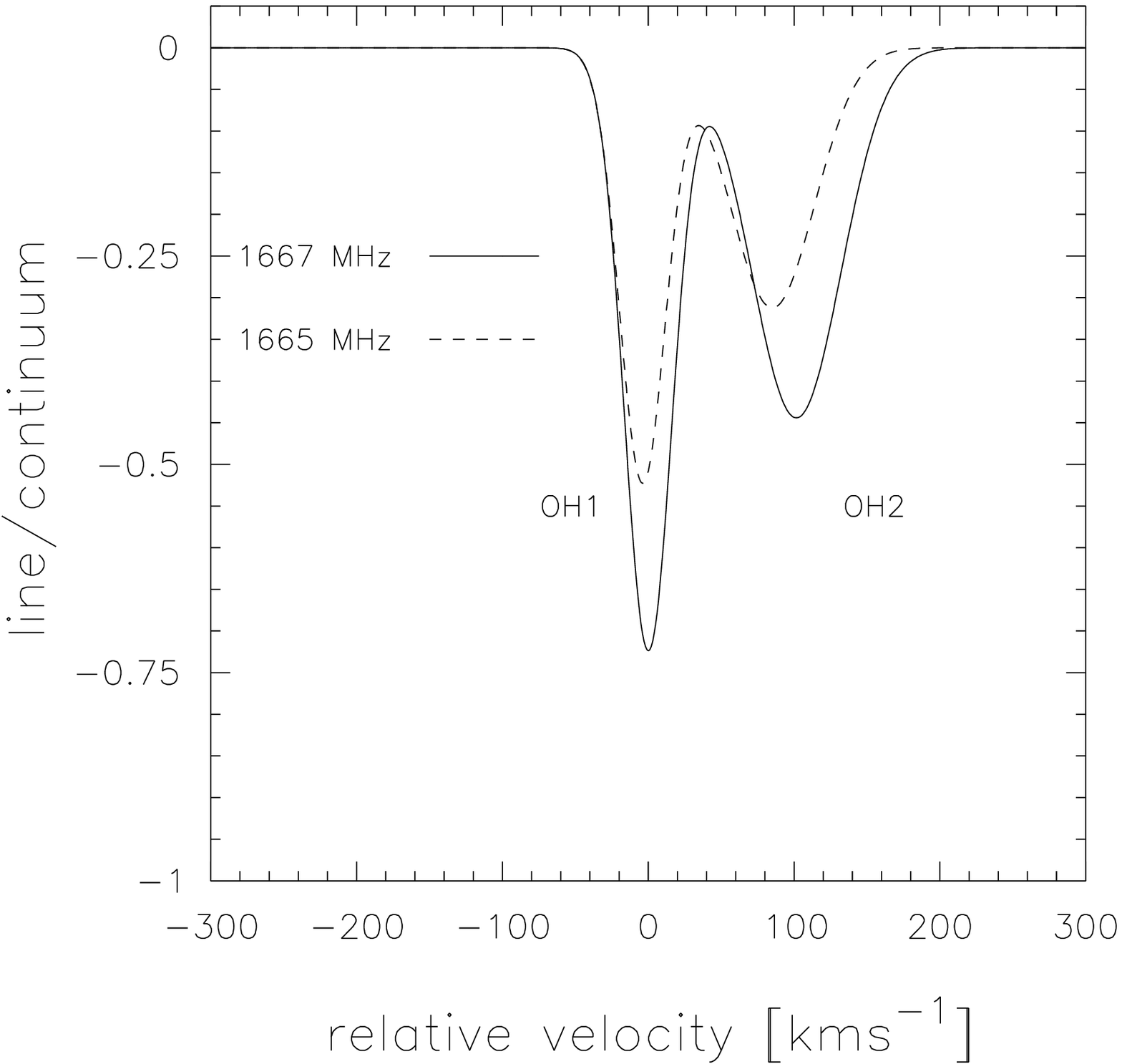}
\caption{$left$: Hydroxyl absorption spectrum of the 1667 MHz and 1665
MHz transitions, obtained with the EVN in November 2000. An rms noise
per channel is 0.7 \mb. The velocity resolution is 11.2
\kmss. Velocities in the spectra are scaled in the radio LSR
convention. The adopted systemic velocity of NGC\,3079 is \vlsr = 1116
\kmss. The peak velocities of \hi\ absorption and \ho maser are denoted by arrows. The results of 2-D Gaussian-fitting of the spectrum are listed
in Table 1. $right$: Gaussian models fitted to the OH spectrum of the
1667 and 1665 MHz transition. The velocity is referenced to \vlsr=
1013 \kmss, the Gaussian-fitted peak velocity of the 1667 MHz OH1 in Table 2.}
\end{figure}
\begin{figure}
\includegraphics[angle=0,scale=0.47]{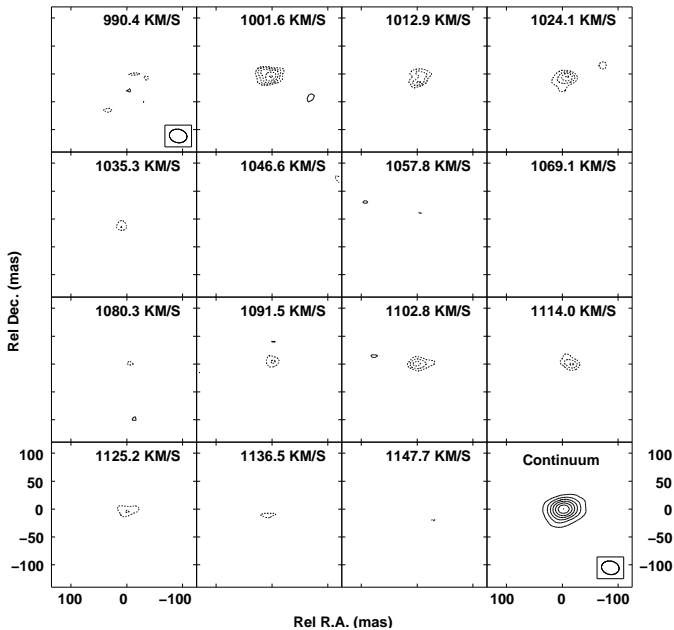}       
\caption{Velocity channel maps of uniformly weighted continuum-subtracted
spectral line cube where the OH absorption at 1667 MHz
is present. The contour levels are -8,-7,-6,-5,-4,-3, 3 $\times$ 0.8 \mb (1 
$\sigma$). 18-cm uniformly weighted continuum map is also shown with the 
 contour levels at -5,5,10,15,20,25,30,35,40 $\times$ 0.25 \mb (1 $\sigma$). 
The center (0,0) position is R.A.(J2000)=10$^h$01$^m$57$^s$.805 
Dec.(J2000)=+55$^{\circ}$40'47''.08. The synthesized beams are plotted at the 
top-left and the bottom-right corner.}
\end{figure}
\begin{figure}
\includegraphics[angle=0,scale=0.45]{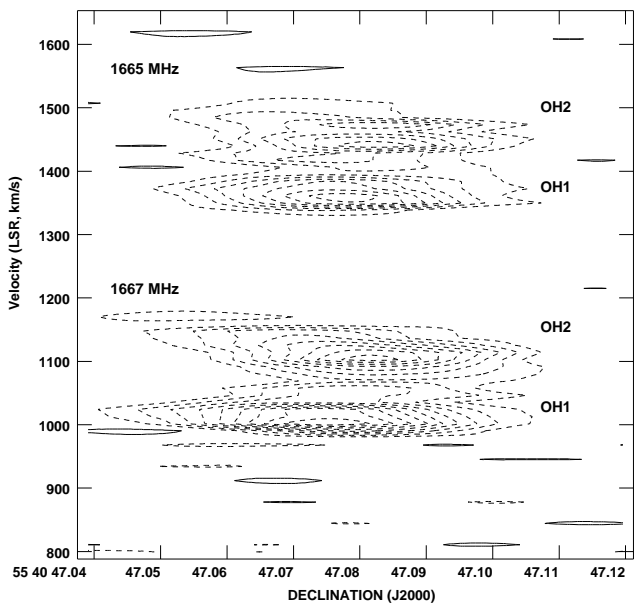}    
\caption{Position-velocity (PV) diagram produced from the naturally weighted 
spectral-line visibility cube, sliced along P.A.= 15 degrees to align the major axis of the CO rotation  \citep{sofu92} and the cuts used in PV diagrams in (Baan \& Irwin 1995). The contours are plotted at 10, 20, 30, 40, 50, 
60, 70, 80, 90, 100 percent of the peak intensity of -6 \mb.
The positive contour corresponds to +0.6 \mb.}
\end{figure}
\begin{figure}
\includegraphics[angle=0,scale=0.45]{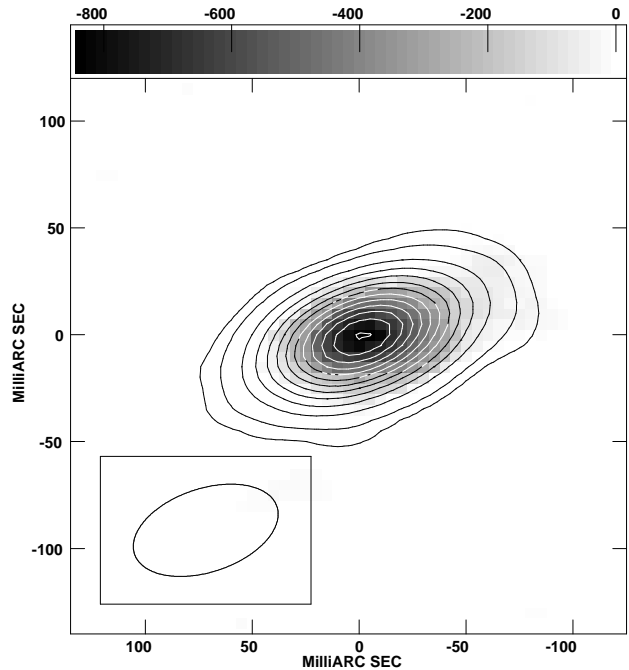}    
\caption{18-cm naturally weighted continuum map (contour) superposed with a
1667 MHz OH intensity (0th moment) map (grey-scaled in \jb~m~s$^{-1}$). The peak flux density of the continuum is 10.3 \mb. The contours are plotted at -5,5,10,20,30,40,50,60,70,80,90,100,110,120 $\times$ 0.085 \mb (1 $\sigma$). The synthesized beam (70 mas $\times$ 39 mas, P.A.= --72$\degr$) is plotted on the lower left corner.}
\end{figure}
\begin{figure}
\includegraphics[angle=0,scale=0.55]{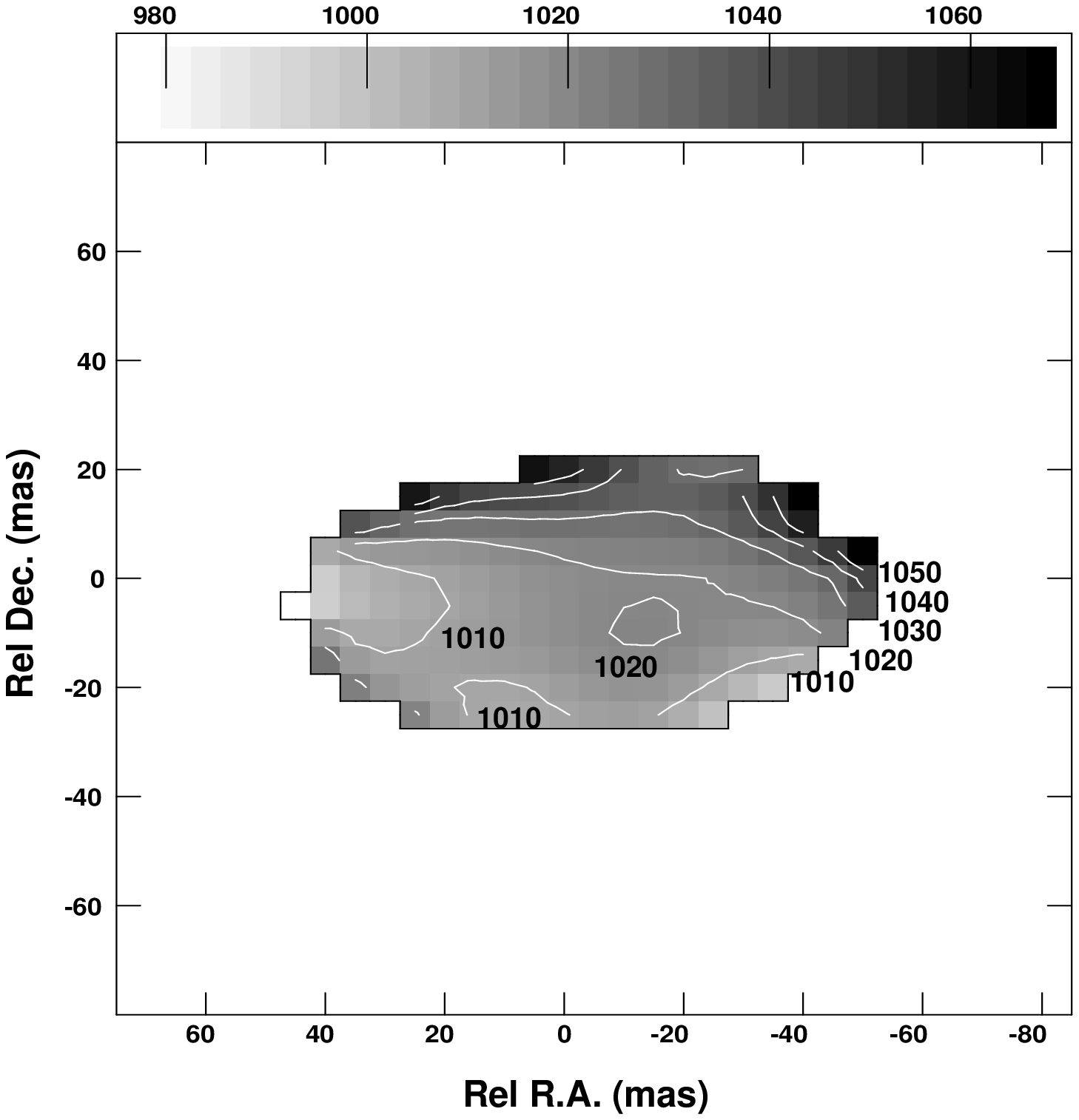}
\includegraphics[angle=0,scale=0.55]{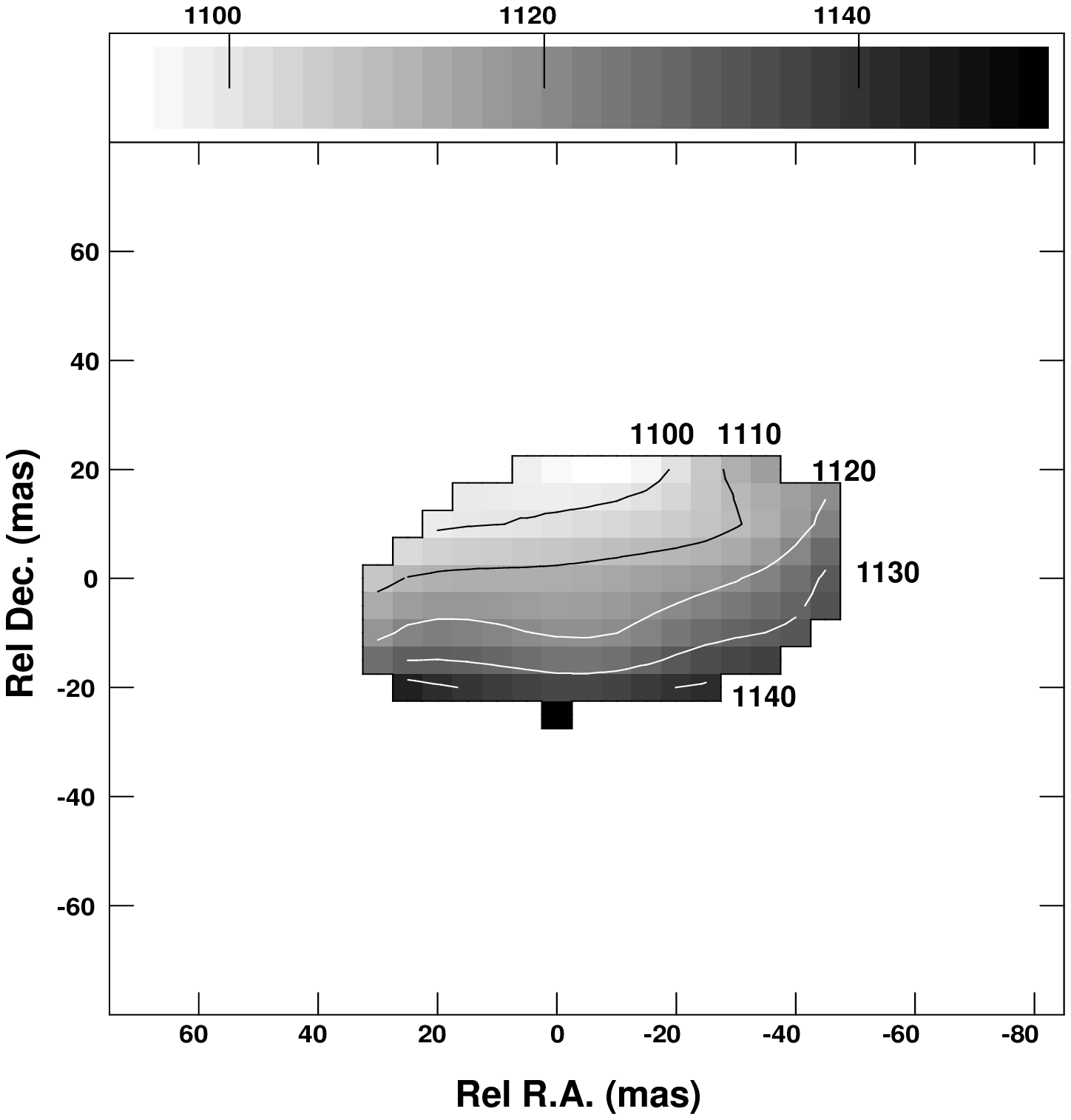}   
\caption{Mean velocity field (1st moment) of the two 1667 MHz OH absorption components. 
Contours are plotted every 10 \kms in LSR velocities with the gray scales 
indicated on the top.
$left$: blue-shifted component (OH1), $right$: systemic velocity component 
(OH2).}
\end{figure}
\begin{figure}
\includegraphics[angle=0,height=9cm,width=9cm]{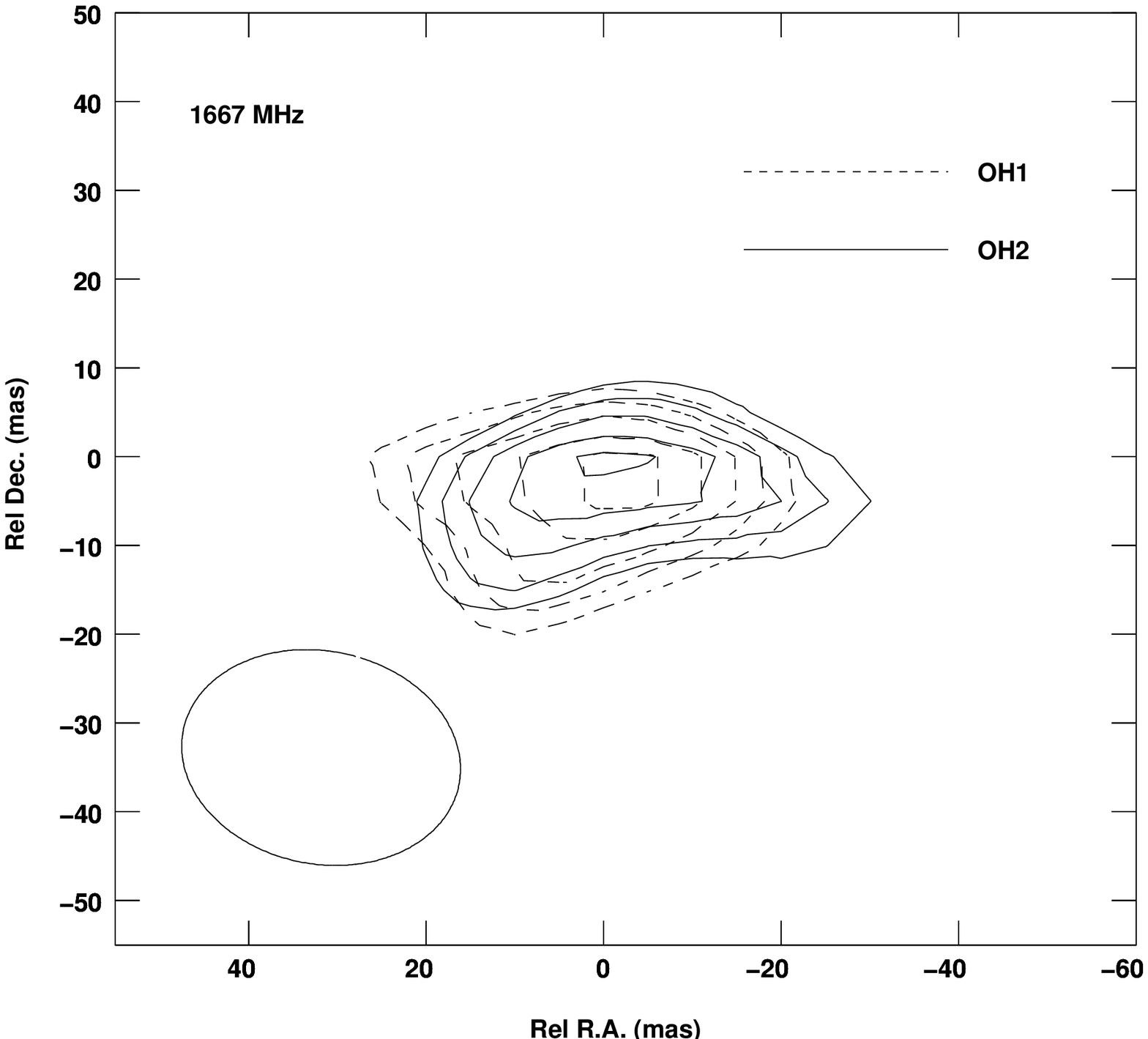}
\includegraphics[angle=0,height=9cm,width=9cm]{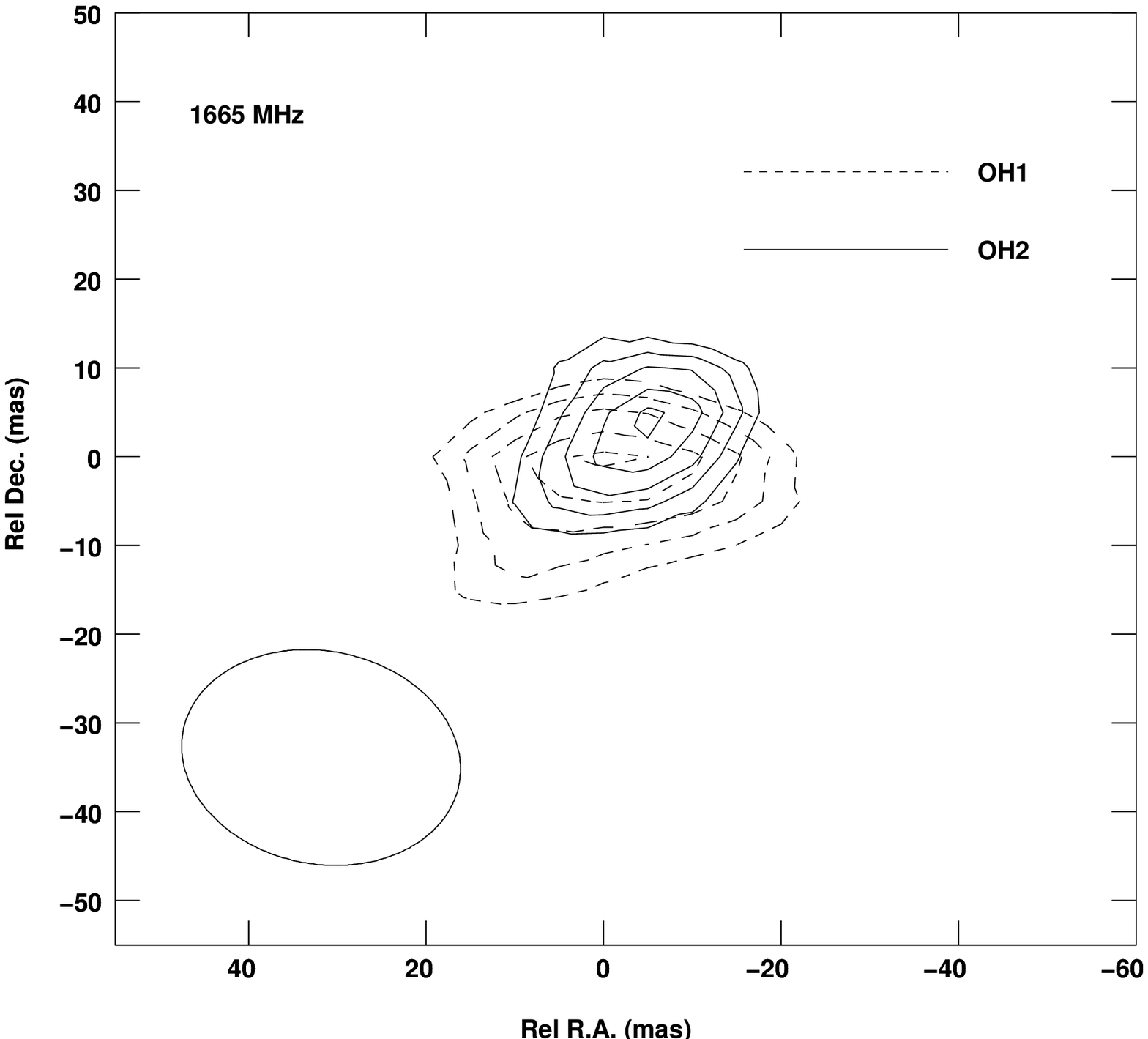}   
\caption{Comparison of the intensity distribution of the OH double 
peaks OH1 and OH2, integrated over 16 velocity channels (corresponding to $\sim$ 180 \kmss)  for each peak. Contour levels run from -2 to -1.2 mJy per beam
in steps of 0.2 mJy per beam.
The synthesized beam is included at the left corner of each plot. 
$left$: 1667 MHz, $right$: 1665 MHz. The dashed and solid lines show OH1 
(blue-shifted absorption) and OH2 (near systemic velocity), respectively.}
\end{figure}
\section[]{Discussion}
\subsection[]{The puzzle of the nuclear radio continuum}
NGC\,3079 is known to exhibit a complex radio structure in the nuclear region. 
Recent VLBI studies reveal a multiple milli-arcsec-scale  nuclear 
source structure with five frequency dependent components \citep{enno03}. The 
continuum components A, B and C are only present in maps at 22 GHz but are 
totally obscured at L-band.  The identification of a true nucleus in the galaxy 
has been a subject to debate.  If the \ho maser structure is indeed confirmed 
to be a disk  with a rotation axis misaligned by 15--20$\degr$  with respect to the kpc-scale CO(1--0) disk at P.A.= 30$\degr$, then the nucleus would lie close to the A-C-B line (Fig. 7). \cite{trot98} proposed that the dynamical center lies 
between components A and B, more specifically at the intersection of the jet 
axis and a north-south distribution of the \ho masers. The north-south 
distribution of the \ho maser was more convincingly confirmed in \cite{kond03} 
by measuring the positions of the new red-shifted \ho maser features reported 
in \cite{hagi02a}. An \ho disk centered close to C and having a systemic 
velocity of 1120 \kms would be consistent with the \ho components at 1123 \kms 
at C, the systemic velocity of the \hi absorption of 1116 \kms, and the 
systemic OH component (OH2) at 1114 \kmss. The models based on the velocity signature of the 
OH and HI absorption \citep{baan95} and of the \hi VLBI signature \citep{sato00}  refer to components E and F and do not reflect the location 
of the nucleus. Various monitoring programs have been executed to detect a 
velocity drift of the systemic components as seen in NGC\,4258, but they have 
failed to find any drifting component to confirm the disk signature (e.g. Baan \& Haschick 1996; Hagiwara et al. 2002a).  \\
 \\
Considering the available evidence,  the true nucleus is hidden at L-band and  
lies close to C on the line connecting components C and B. The nature of the L-band components  E and F  is not yet certain.  These components could be 
circum-nuclear starburst regions but for that E has too high a brightness 
temperature and could be a supernova remnant.
\subsection[]{The nature of the OH absorption}
An important result of our EVN observations is the interpretation of the two 
different velocity fields traced by the two OH absorption components on 
scales of 10 mas towards the unresolved radio continuum nucleus.  The spectrum 
in Fig. 1 shows that most of the OH absorption has been  resolved out in our 
data set, as compared with single-dish and VLA spectra in \cite{baan95}. This 
implies that we observe compact gas that extends no more than the beam size of 
45 mas, or $\sim$ 4 pc, which is almost equal to the size of the background 
continuum. The OH absorbing gas has been interpreted primarily to be a part of 
circum-nuclear gas traced by the CO(1--0) emission \citep{baan95}. A single black body fit based on the measurement of far-infrared 
(FIR) flux densities at 12 -- 100 $\micron$ in NGC\,3079 gives a temperature of 43 K and the maximum extent of 
this large-scale FIR source (e.g. torus) of about 130 pc. Our EVN data do not 
provide  compelling evidence for such a large-scale torus.
\subsubsection[]{The absorbing gas structure}
The nucleus of NGC\,3079 hosts a LINER or a type 2 Seyfert
nucleus. Therefore, the nucleus could be obscured by an edge-on dusty
torus or intervening medium along the line of sight. Because of the
systematic similarity of the \hi and OH absorptions, \cite{baan95}
suggested a possible association of the two \hi and OH absorbing gas
components with the double continuum source (Fig. 7).  With a 1.0
arcsec (corresponding to 80 pc) resolution of the VLA in A
Configuration, the data in \citet{baan95} were insufficient to extract
precise connections between these absorption components and the
nuclear radio continuum structure. \\
Interpretation of the VLBI data in \citet{sato00} shows that the three
\hi components are resolved and that the column densities of each
component against F (blue-shifted \hii), F+E (systemic \hii), and E
(red-shifted \hii) are almost the same, suggesting that E and F are
uniformly obscured by the ISM foreground to the nuclear sources.  Our
EVN data failed to resolve E from F, so that we cannot identify the
individual contributions for OH absorption and the OH column
densities for each component with those of \hii.  \citet{baan95}
introduced a rotating disk model, in which the torus is confined in
the roughly north-south orientation with an inclination about
--20$\degr$ w.r.t the CO(1--0) disk standing at P.A.= 15$\degr$. In this
discussion, the jet axis connecting the nuclear radio sources of E and F
makes an angle of about 45$\degr$ with the confining torus.  The sense
of the rotation of the torus proposed in \citet{baan95} is consistent
with that of the CO disk/core and the edge-on \ho maser disk proposed
in \citet{trot98} and \citet{kond03}. \citet{sato00} proposed a
counter-rotating disk at P.A. = 30$\degr$ in order to explain the
spatially resolved \hi absorption. However, the directions and
position angles of the torus and the CO and \ho maser disks are completely
inconsistent with those of the torus model in \cite{sato00} and the
evidence for this counter-rotation is weak.  \\

On the other hand, a counter-rotating structure  reversed to the 
kiloparsec-scale CO(1--0) disk has  been observed in OH absorption in a type 2 
Seyfert galaxy  NGC\,5793 \citep{hagi00}. Similarly, two nuclear disks with 
radii $\sim$ 100 pc embedded in the outer kpc-scale gas disk have been clearly 
resolved in CO(2--1) in the merging nuclei of the ultra-luminous infrared galaxy 
Arp\,220 (IC\,4553). The rotational sense between these two disks is reversed 
due to the counter rotation of the two nuclei themselves \citep{kazu99}.
\subsubsection[]{Interpretations of the double OH peaks}

The OH absorption in the EVN spectrum shows double peaks with a separation of 
101.6 \kms in velocity (Table 2). \cite{baan95} considered the association of 
OH1 and OH2 with the radio twin peaks E and F (Fig. 7) with an OH velocity 
gradient between the two components along the jet axis of approximately 50 cos 
$\phi$ \kms pc$^{-1}$. The projected separation of the twin radio peaks E and F 
 is $\sim$ 25 mas or 2 pc and the projected angle ($\phi$) lies between the 
radio axis E--F and the plane of the OH gas motion. The value of the gradient 
would be 10 times larger than that of the CO(1--0) kpc-scale disk and the 
CO(1--0) nuclear core rotating in the north-south direction, unless cos $\phi$ 
is  unrealistically small.  Consequently, it is not plausible to correlate the 
OH velocity difference between E and F and the internal 
velocity gradients of the CO(1--0) disk/core, although the direction of the OH 
velocity gradients agrees with that of the larger-scale CO(1--0) disk. 
This velocity gradient would also be about 10 times larger than those seen in 
the merging galaxy Arp\,220 \citep{mund01} and the Seyfert galaxies Mrk\,231 
and Mrk\,273 \citep{hans03a,hans03b}, where HI absorption or OH maser emission 
reveal a rotating molecular torus with an inner radius of several tens of 
parsecs from the central engine. \\
Considering the available evidence, the double peaks must arise from two 
kinematically independent systems in the nuclear region. The direction of the 
velocity gradient of the systemic OH2 is in good agreement with that of the CO 
and \ho maser disks. In addition, the velocity range of OH2 nearly coincides 
with the systemic velocity of NGC\,3079. We know that the \hi absorption at the 
systemic velocity is seen against the whole nuclear continuum at 21-cm 
\citep{sato00}. This can be accounted for by clumpy gas in a kiloparsec-scale 
disk.  Accordingly, the systemic OH2  probes gaseous components in the CO disk, 
and particularly the  inner disk on scales of 10-100 pc. \\
On the other hand, the velocity field of the blue-shifted OH1 is very different 
from that of OH2 and of the CO disk. \citet{baan95} argued that a foreground 
and an expanding shell driven by the nuclear superbubble could explain the 
large blue-shift of the OH1 centroid velocity, while a receding shell may 
account for the other weak red-shifted OH absorption component around \vlsr = 
1260 \kmss, as observed with VLA. In the starburst galaxy M\,82, distinct 
outflow components of molecular gas were discovered that extend over 500 pc 
above the plane of the disk  and lie along the minor axis of the galaxy (e.g. 
Nakai et al. 1987). \\
It is uncertain whether or not the outflows traced by OH1 are
associated with the (possible) starburst-related components E and F themselves because the velocity gradient of OH1 does not align with this E--F axis (Fig. 7).  It should be noted
that the velocity range of OH1 (\vlsr = 956--1050 \kmss) overlaps with
that of several blue-shifted \ho features peaking at velocities \vlsr
= 1012, 1018, 1034, and 1035 \kms \citep{trot98}.  Although these
features have a velocity signature similar to that of the OH1
absorption component, there is no evidence for a physical
association. Some of the \ho maser components outside the 
edge-on masering disk of the Circinus galaxy are interpreted as molecular
outflow components, ejected from the edge-on structure \citep{linc03}. \\
The two OH emission features identified at \vlsr= 910 \kms and 1230 \kms in 
\cite{baan95} and detected weakly in our data may be accounted for by the 
blue-shifted 1667 MHz and 1665 MHz OH features that lie at velocities just 
below the OH1 outflow components. They may arise in the shocked molecular 
components foreground to the outflow structure.   In the above picture, an 
association with molecular outflows can explain the blue-shifted OH1, the weak 
OH emission, and also possibly the \ho maser components.  However, this last association cannot be quantified due to difficulties in comparing the distribution of weak \ho masers and the  OH absorption  on very different angular scales.
\section[]{Conclusions and Summary}

The broad OH absorption towards the nuclear continuum source in NGC\,3079 was 
imaged using VLBI techniques. The EVN observations reveal two kinematically independent OH 
absorption components,  where OH1  is the blue-shifted absorption, and OH2  is 
the absorption at the systemic velocity of the galaxy.  
The understanding of the kinematics of these components is limited by the fact that the 18-cm background continuum and the OH components have not been spatially resolved. 

  \noindent 1. The OH2 component shows a distinct velocity gradient in roughly 
north-to-south direction viewed against the unresolved nuclear continuum at 
18-cm consisting of components E and F. This velocity gradient of the systemic 
OH2 is consistent with that of the clumpy gas component in the kpc-scale CO 
disk (P.A.= 15$\degr$).    

\noindent 2. The blue-shifted component OH1 appears to be associated
with molecular outflows and is possibly associated with the nuclear
super-bubble due to circum-nuclear starburst activity at components E
and F and possibly at other locations. The velocity gradient of the
OH1 component is not yet understood butis almost reversed from that of
the systemic component OH2. The weak OH emission features marginally
detected in our EVN observation may also be associated with shocked
gas in these nuclear outflows.

\noindent 3. The \ho masers in the galaxy have been attributed to a compact 
disk structure around the true nucleus located west of the radio continuum 
component C. The OH1 velocity range is similar to  that of the blue-shifted
 \ho maser components and there may be some relation. 

The combined knowledge of our OH data, of previous VLBI data on the radio 
continuum, of \ho masers, and of the HI absorption on scales of 100 pc down to 
sub-parsec scales, provides a clear and consistent interpretation of the 
various spectral components seen at the nucleus. 
VLBI observations at even higher angular resolution and high sensitivity could 
resolve the nature of the OH and \hi components in relation to \ho maser 
components in the nuclear region of NGC\,3079. 

\section*{Acknowledgments}
This research has made use of the NASA/IPAC Extragalactic Database
(NED), which is operated by the Jet Propulsion Laboratory, California
Institute of Technology, under contract with the National Aeronautics
and Space Administration. The European VLBI Network is a joint
facility of European, Chinese, South African and other radio astronomy
institutes funded by their national research councils.  We thank the
anonymous referee for useful comments.
%
%
\bibliographystyle{mn2e}
\bibliography{papers}
\bsp

\onecolumn
\begin{figure}
\includegraphics[angle=0, scale=0.65]{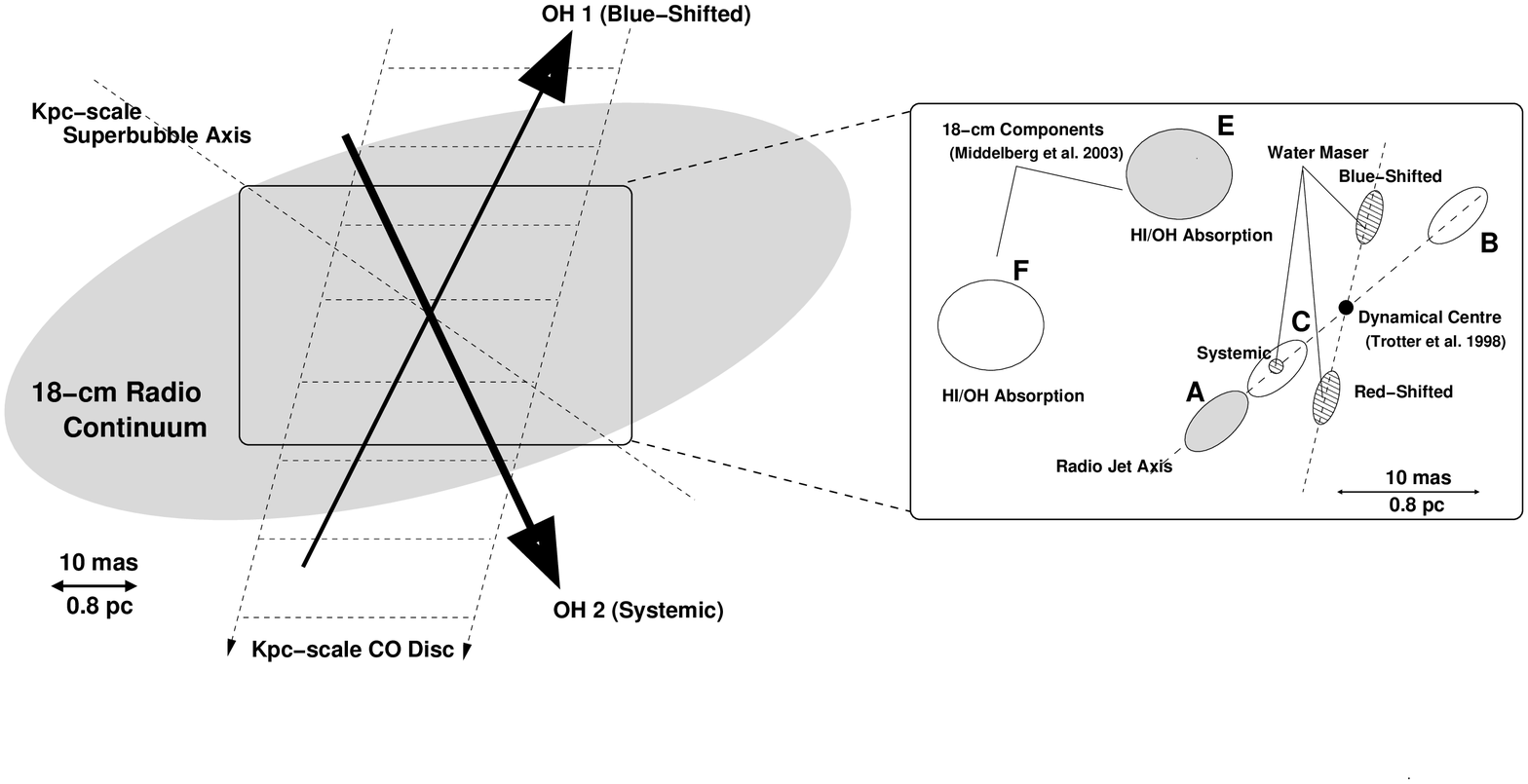}  
\caption{Schematic views of the nuclear region of NGC 3079. The sizes
of components are drawn roughly in scale. Oval indicates the outline
of the unresolved continuum nucleus in Fig. 2. Upward thin arrow
denotes the direction of OH1 velocity gradient against E and F, both
of them are not resolved in our EVN observations. Downward thick arrow shows
the sense of the OH2 velocity gradient that is roughly consistent with
the kpc-scale CO rotation, indicated by thin dashed arrows. A thin
dashed line denotes the axis of the kiloparsec-scale super-bubble
consisting of wide-angle outflows ejected from the nuclear
region  \citep{ceci01}. Inset displays the innermost nuclear region of NGC\,3079,
where three radio sources A, B, and C are aligned along P.A.= 55$\degr$.  
The weak component C may be shaded by the inclined torus or
disk traced by the \ho masers \citep{trot98}.  Components E and F are
visible at lower frequencies at from 1.7 to 5.0 GHz
\citep{kond03,enno03} and may result from circum-nuclear
starburst. The component labels are adopted from the nomenclature
defined in \citet{irwi88} and \citet{enno03}. Components E and A show
similar spectral characteristics, that are different from other components.
The spectra of E and A are inverted at 2.3 and 5.0 GHz, respectively \citep{enno03}.}
\end{figure}
\begin{table}
\caption{Gaussian-fitted parameters of the OH absorption and emission spectra. VLA data are from Baan \& Irwin (1995). Vc refers a line-peak velocity. Sp and $\Delta$V are peak flux density and linewidth (FWHM), respectively. The EVN measurements of the OH emission are tentative.}
\label{tbl-2}
\begin{tabular}{lllllll}
\hline
& \multicolumn{3}{c}{VLA} & \multicolumn{3}{c}{EVN (This paper)} \\ 
%

&Vc&Sp& $\Delta$V&Vc&Sp&$\Delta$V \\
&(\kmss)&(\mb)&(\kmss)&(\kmss)&(\mb)&(\kmss)    \\
\hline
OH ABSORPTION & \\
~~OH1  (1667 MHz)  & 1013 &22&84& 1011.9 $\pm$ 0.9&  8.5 $\pm$ 0.4 & 38.7 $\pm$ 2.2 \\
~~~~~~~~~(1665 MHz)  & 1354     &15& ~- & 1361.6 $\pm$ 1.3 & 6.1 $\pm$ 0.4 &  37.2 $\pm$ 3.1   \\

~~OH2  (1667 MHz)  & 1114  &27&92&  1113.5 $\pm$ 2.0 & 5.2 $\pm$ 0.3 & 72.1 $\pm$  5.0\\
~~~~~~~~~(1665 MHz)  & 1462       &18& ~- &  1450.7 $\pm$ 2.9 & 3.6 $\pm$ 0.3 & 67.8 $\pm$ 7.3\\

OH EMISSION & \\
~~~~~~~~~(1667 MHz)  & ~910 & 4.7  & 61 & ~($\sim$900)  & (0.5)  & ~~~~~-\\
~~~~~~~~~(1665 MHz)  & 1260 & 1.7  & ~-  & ($\sim$1235)  & (0.5) & ~~~~~-\\
%
\hline
\end{tabular}
\end{table}
\begin{table}
\caption{Parameters of the 1667 MHz OH absorption. Global-VLBI data are from 
Sawada-Satoh et al. (2001) and VLA data are from Baan \& Irwin (1995).}
\label{tbl-3}
\begin{tabular}{lllllll}
\hline
& 
\multicolumn{2}{c}{EVN}&\multicolumn{2}{c}{Global-VLBI}&\multicolumn{2}{c}{VLA 
(A Configuration)} \\
%
%
&~~OH1 (Blue-shift)  & OH2 (Systemic)&~~OH1  & OH2 & OH1 & OH2\\
\hline

Optical depth ($\tau$)  &  0.46 $\pm$ 0.03 &  0.31 $\pm$ 0.02 &  0.73 $\pm$ 0.17   & 0.79 $\pm$ 0.13 
& 1.03 & 1.28 \\
N$_{\rm OH} / {\rm T_{\rm ex}}$ (K $^{-1}$ cm$^{-2}$) & 4.3 $\times$
10$^{15}$ & 5.3 $\times$ 10$^{15}$ & 9.2 $\times$ 10$^{16}$ & 1.8
$\times$ 10$^{17}$& 2.2 $\times$ 10$^{16}$ & 2.5 $\times$ 10$^{16}$ \\

\hline
\end{tabular}
\end{table}

\end{document}